\documentclass[twocolumn,aps,prd]{revtex4}

\usepackage{mathrsfs}
\usepackage{amsfonts}
\usepackage{graphicx}%
\usepackage{dcolumn}
\usepackage{amsmath}

\newcommand{\be}{\begin{equation}}
\newcommand{\ee}{\end{equation}}
\newcommand{\bey}{\begin{eqnarray}}
\newcommand{\eey}{\end{eqnarray}}
\newcommand{\ba}{\begin{array}}
\newcommand{\ea}{\end{array}}
\newcommand{\bi}{\begin{itemize}}
\newcommand{\ei}{\end{itemize}}
\newcommand{\bem}{\begin{enumerate}}
\newcommand{\eem}{\end{enumerate}}
\newcommand{\bw}{\begin{widetext}}
\newcommand{\ew}{\end{widetext}}

\newcommand{\ra}{\rangle}
\newcommand{\la}{\langle}
\newcommand{\pp}{\partial}
\newcommand{\ov}{\overline}

\newcommand{\ww}{\widetilde}

\newcommand{\bk}{{\bf k}}
\newcommand{\bp}{{\bf p}}

\newcommand{\bq}{{\bf q}}

\newcommand{\E}{{\cal E}}

\newcommand{\HH}{{\mathscr{H}}}

\newcommand{\N}{{\mathscr{N}}}

\begin{document}

 \title{Neutrino-mass differences studied in a model with one basic neutrino
 }

\author{Wen-ge Wang \footnote{ Email address: wgwang@ustc.edu.cn}}
\affiliation{
 Department of Modern Physics, University of Science and Technology of China,
 Hefei 230026, China
 }

 \date{\today}

 \begin{abstract}
 In this paper, we propose and study a model, in which there exists only one basic neutrino.
 Basically, the Lagrangian of the model is
 obtained from the Lagrangian in the Glashow-Weinberg-Salam electroweak theory
 by reducing the three flavor neutrinos there to one basic neutrino.
 In this model, neutrino states with a fixed flavor are interpreted as
 certain superpositions of states of the basic neutrino and states of the related charged lepton.
 Neutrino mass states are associated with low-lying eigenstates of the total Hamiltonian.
 We derive an approximate expression for a ratio of neutrino-mass differences,
 which gives a value $17$ for the ratio, about half of the experimental result of $33$.
 \end{abstract}


 \maketitle

\section{Introduction}

 Neutrino oscillation
 \cite{Fukuda98,Ahmad02,Araki05,Adamson08,An12,Ahn12,Abe12,Abe14,GN03,SV06}
 is one of the most important discoveries made in the past two decades in physics.
 This phenomenon can not be described within the standard model (SM),
 the most successful quantum field theory that
 has ever been developed up to now \cite{Weinberg-book,Peskin}.
 To give a proper explanation to it is still a big challenge to theoretical physics
 and this is one of the reasons for the topic of going beyond the SM to
 attract lots of attentions (see, e.g., \cite{book-fundam-QFT,GGS99,KO01}).

 The neutrino oscillation may be described in a simple way,
 if the three flavor neutrinos observed experimentally,
 namely, the electron neutrino, muon neutrino, and tauon neutrino,
 are assumed to be mixtures of three mass states of neutrino.
 Based on this understanding,
 the simplest way of going beyond the SM at a phenomenological level is to insert
 some three-dimensional mass matrix into the total Lagrangian in the SM,
 given in the basis of the three flavor neutrinos (see, e.g., the review paper Ref.\cite{SV06}).
 Due to the phenomenological nature of this treatment, it is inevitable for the mass matrix to have many
 free parameters, although arguments based on symmetry
 consideration may suggest some relations among them.

 Going further along the above-discussed line of understanding neutrino oscillation,
 an interesting question is whether the three neutrino mass states may be studied as eigenstates
 of some total Hamiltonian in a very large state space,
 not restricted within a presumed, three-dimensional subspace spanned by flavor neutrino states.
 In such an approach,
 an interpretation of flavor neutrino states would be needed within the large state space.
 An idea to be adopted in this paper is that, if three types of states can be found in such a large state space,
 all of which have certain relation to neutrino, such that each type
 of them interacts effectively with one species of charged lepton only, or almost so, 
 then, these three types of states may be interpreted as three flavor neutrino states.
 If such states may be found, then,
 there would be no compelling reason to stick to the assumption of three species of neutrino
 at the fundamental level;
 in other words, the possibility that there may exist only one basic neutrino at the fundamental level
 should not be excluded.

 In this paper, we propose and study a model, in which there exists only one basic neutrino
 at the fundamental level.
 Except for this point, basically, the model is constructed in a way similar to
 the Glashow-Weinberg-Salam (GWS) electroweak theory \cite{Weinberg-book,Peskin}.
 In this model, we find three types of states
 that can be interpreted as flavor neutrino states in the sense discussed above.
 The three neutrino mass states are associated with
 low-lying eigenstates of the total Hamiltonian.
 We derive an approximate expression for the three neutrino-mass states,
 which includes an unknown parameter related to the basic neutrino.
 But, the neutrino-mass differences predicted by the expression, which are mainly due to
 interactions among leptons, do not depend on the unknown parameter; even further, 
 their ratios can be approximately expressed in terms of experimentally-measured
 masses of charged leptons.

 The paper is organized as follows.
 Basic contents of the proposed model are discussed in Sec.\ref{sect-model}.
 In Sec.\ref{sect-e-neutrino}, we discuss properties of some subspaces of the total state space
 and show that ground states of the parts of the total Hamiltonian within these subspaces
 may be associated with flavor neutrino states.
 Then,  in Sec.\ref{sect-mass-state},
 we derive an approximate expression for neutrino masses
 and compare some of its predictions with experimental results.
 Finally, conclusions and discussions are given in Sec.\ref{sect-conclusion}.

\section{The proposed model}\label{sect-model}

 In this section, we introduce the proposed model and discuss its basic properties.

\subsection{Basic contents of the model}

 In the proposed model, we consider only leptons and
 vector bosons mediating the electroweak interactions.
 There are eight elementary particles: one basic neutrino, three charged leptons
 (electron, muon, and tauon), and four vector bosons (photon and bosons $Z^0$ and $W^\pm$).
 We assume that the basic neutrino can not be experimentally detected;
 what are detected in experiments are flavor neutrino states,
 the descriptions of which will be discussed later.

 We use $\nu$ to indicate the basic neutrino
 and use $\kappa =e, \mu$, and $\tau$ to indicate the three charged leptons
 of electron, muon, and tauon, respectively.
 The corresponding antiparticles are denoted by $\ov\nu$ and $\kappa^+$, respectively.
 The left-handed fields of the basic neutrino $\nu$ and of
 a charged leptons $\kappa$ are written in a doublet form,
\begin{gather}\label{}
 E_L^\kappa =  \left(\begin{array}{l} \nu \\ \kappa \end{array} \right)_L,
 \quad \kappa =e, \mu, \nu.
\end{gather}
 The right-handed fields of $\kappa$ are indicated by $\kappa_R$.
 Ignoring terms related to masses, the Lagrangian ${\cal L}$ for the eight particles
 in the proposed model has a form, which is obtained from the Lagrangian in the GWS theory
 by reducing the three flavor neutrinos there to one basic neutrino.
 Specifically, it is written as
\begin{gather}\label{}\notag
 {\cal L} = \sum_\kappa \ov E_L^{\kappa} (i \gamma^\mu \pp_\mu) E_L^\kappa + \ov\kappa_R (i \gamma^\mu \pp_\mu) \kappa_R
 + e A_\mu J^{\mu\kappa}_{EM}
 \\ + g(W^+_\mu J^{\mu\kappa+}_W + W^-_\mu J^{\mu\kappa-}_W + Z^0_\mu J^{\mu\kappa}_Z)
 + {\cal L}_B, \label{Lag}
\end{gather}
 where ${\cal L}_B$ represents the Lagrangian for the bosons and
\begin{gather}\label{}\notag
 J^{\mu\kappa+}_W = \frac{1}{\sqrt 2} \ov \nu_L \gamma^\mu \kappa_L,
 \quad J^{\mu\kappa-}_W = \frac{1}{\sqrt 2} \ov \kappa_L \gamma^\mu \nu_L,
 \\ \notag J^{\mu\kappa}_Z = \frac{1}{\cos \theta_w} \Big[ \ov\nu_L\gamma^\mu \Big(\frac 12\Big) \nu_L \hspace{3cm}
 \\ \hspace{1cm} + \ov\kappa_L \gamma^\mu \Big(-\frac 12 + \sin^2\theta_w \Big)\kappa_L \notag
  + \ov\kappa_R \gamma^\mu (\sin^2\theta_w)\kappa_R \Big], \notag
 \\ J^{\mu\kappa}_{EM} = \ov\kappa \gamma^\mu(-1)\kappa. \hspace{4cm}
\end{gather}

 Bare masses of the three charged leptons and of the bosons  $Z^0$ and $W^\pm$
 can be introduced in the same way as those in the GWS theory,
 by the Higgs mechanism with the Higgs field.
 Similarly, a bare mass can be given to the basic neutrino,
 if one assumes that it is a Dirac fermion and has a right-handed field.
 We are not to give further discussion for this part of the model, because it will not be used in the discussions
 to be given below for neutrino states.

 From the form of the Lagrangian ${\cal L}$ given above, it is seen that,
 for each pair $(\nu, \kappa)$ of $\kappa =e, \mu, \tau$,
 the interaction Lagrangian can be introduced by the gauge symmetry group
 $U(1)\otimes SU(2)$, in the same way as that in the GWS theory
 for the corresponding generation of lepton.
 This gauge symmetry guarantees that the proposed model
 can be renormalized in the same way as the GWS theory.

 Moreover, it is not difficult to verify that the proposed model has the same anomalies
 of Feynman diagrams as the GWS theory.
 As is known, owing to the gauge symmetry in the theory, the anomalies in the GWS theory
 can be cancelled within each generation separately,
 when contributions from quarks are taken into account.
 Then, since each pair $(\nu, \kappa)$ in the proposed model has the same gauge symmetry as
 the corresponding generation of leptons in the GWS theory,
 anomalies in the proposed model can also be cancelled when contributions from quarks are taken into account.

 Finally, we note that, since the basic neutrino field appears in each of the three doublets $E_L^\kappa$,
 if there is no restriction to rotations in the spaces of the three
 doublets  with respect to the SU(2) symmetry,
 a mixture of the charged leptons would be allowed, which has never been observed experimentally.
 In order to avoid such a mixture, some restriction must be imposed to the rotations.
 For example, one may assume that certain quantum number
 (conserved quantity) may be assigned to the three charged leptons,
 such that rotations are  forbidden within the three spaces,
 or allowed to happen within only one of the three spaces.
 We are not to give further discussions for this point,
 because this is irrelevant to the main purpose of this paper.

\subsection{Effective Hamiltonian}

 As discussed above, the proposed model is renormalizable.
 We assume that, in a renormalized version of the model, the total Hamiltonian has eigenstates with
 finite eigenvalues at least in a low energy region.
 In the following part of this paper, we discuss within such a renormalized
 version of the model.

 The state space of the renormalized model is spanned by Fock states,
 which are eigenstates of a renormalized free Hamiltonian denoted by $H_0$,
\begin{gather}\label{}\notag
 H_0 =  \int  d\ww p \sum_{\alpha,F} \epsilon_{F \bp}  \left(b^{\alpha \dag}_{F \bp} b^{\alpha}_{F \bp}
 + d^{\alpha \dag}_{F \bp} d^{\alpha}_{F \bp} \right)
 \\ + \sum_{\lambda,B}  \epsilon_{B \bp} a^{\lambda \dag}_{B \bp} a^{\lambda}_{B \bp},
\end{gather}
 where $\bp$ represents the $3$-momentum, $\alpha$ and $\lambda$ label the spin,
 $d\ww p = \frac{1}{p^0} d^3p$, and $\epsilon_{F \bp}$ and $\epsilon_{B \bp}$
 are kinetic energies of free particles, with renormalized masses $\epsilon_{F {\bf 0}}$
 and $\epsilon_{B {\bf 0}}$, respectively.
 Here, $F$ indicates free fermions, with $F=\nu, e, \mu$, and $\tau$,
 and $B=Z, W^\pm$, and $A$ for the $Z^0$-boson,
 $W^\pm$-bosons, and photon, respectively;
 $b^{\alpha \dag}_{F \bp}$ and $d^{\alpha \dag}_{F \bp}$ are creation operators for the fermions
 $F$ and their antiparticles, respectively, and $a^{\lambda \dag}_{B \bp}$ are creation operators for the bosons $B$;
 the corresponding annihilation operators are written as
 $b^{\alpha}_{F \bp}, d^{\alpha}_{F \bp}$, and $a^{\lambda}_{B \bp}$, respectively.
 The kinetic energies satisfy the ordinary relation,
 $\epsilon_{F(B) \bp}^2= \epsilon_{F(B) {\bf 0}}^2 +\bp^2$.

 We use $H$ to denote the renormalized total Hamiltonian and write it as
\begin{gather}\label{H-total}
  H= H_{0}+ V,
\end{gather}
 where $V$ indicates the renormalized interaction Hamiltonian, the form of which can be obtained from
 the bare Lagrangian in Eq.(\ref{Lag}).
 It is written as
\begin{gather}\label{V-total}
 V = V_{ Z} + \sum_\kappa V_{\kappa} + V_{B},
\end{gather}
 the meanings of which we explain below in detail.
 The term $V_{Z}$ in Eq.(\ref{V-total}) represents the interaction between the basic neutrino and $Z^0$-boson,
\begin{gather}\label{}\notag
 V_Z = \int  d\ww p d\ww q d\ww k \sum_{\alpha,\beta,\lambda}
 R_{1}^Z \delta^3(\bk+\bq -\bp) a^{\lambda \dag}_{Z \bk} b^{\beta \dag}_{\nu \bq} b^{\alpha}_{\nu \bp}
 \\ + R_{2}^Z \delta^3(\bk+\bq -\bp) a^{\lambda \dag}_{Z \bk} d^{\beta \dag}_{\nu \bq} d^{\alpha}_{\nu \bp} \notag
 \\ + R_{3}^Z \delta^3(\bk-\bq -\bp) a^{\lambda \dag}_{Z \bk} d^{\beta}_{\nu \bq} b^{\alpha}_{\nu \bp}  + {\rm c.c.},
 \label{VZ}
\end{gather}
 where $R_{r}^Z$ of $r=1,2,3$ represent the interaction amplitudes.
 Note that, in the interaction $V_Z$, the nonphysical term of creation
 of three particles from the vacuum and its conjugate term have been omitted.

 The interaction term $V_\kappa$ in Eq.(\ref{V-total}) involves the charged lepton $\kappa$ and has the form,
\begin{gather}\label{V-kappa}
  V_\kappa =  V_{\kappa \nu} +  V_{\kappa Z} + V_{\kappa A},
\end{gather}
 where $V_{\kappa \nu}$ represents the interaction between the lepton $\kappa$ and
 the basic neutrino, mediated by $W^\pm$ bosons,
\begin{gather}\label{}\notag
 V_{\kappa \nu} = \int  d\ww p d\ww q d\ww k \sum_{\alpha,\beta,\lambda}
 R_{1}^{\kappa \nu} \delta^3(\bk+\bq -\bp) a^{\lambda \dag}_{W^+ \bk} b^{\beta \dag}_{\kappa \bq} b^{\alpha}_{\nu \bp}
 \\ + R_{2}^{\kappa \nu} \delta^3(\bk+\bq -\bp) a^{\lambda \dag}_{W^- \bk} d^{\beta \dag}_{\kappa \bq} d^{\alpha}_{\nu \bp} \notag
 \\ + R_{3}^{\kappa \nu} \delta^3(\bk -\bq -\bp) a^{\lambda \dag}_{W^+ \bk} d^{\beta }_{\kappa \bq} b^{\alpha}_{\nu \bp}  \notag
 \\ + R_{4}^{\kappa \nu} \delta^3(\bk -\bq -\bp) a^{\lambda \dag}_{W^- \bk} b^{\beta }_{\kappa \bq} d^{\alpha}_{\nu \bp}  + {\rm c.c.}.
\end{gather}
 The interaction amplitudes $R_{r}^{\kappa \nu}$ of $r=1,2,3,4$ are in fact independent of the
 specific species of $\kappa$.
 The term $V_{\kappa Z}$ in Eq.(\ref{V-kappa}) represents the interaction between the charged lepton
 $\kappa$ and $Z^0$-boson, which has a form similar to $V_Z$ in Eq.(\ref{VZ}).
 The term $V_{\kappa A}$ represents the interaction between the lepton $\kappa$ and photon
 and it can also be written in a form like that   in Eq.(\ref{VZ}).

 Finally, the term $V_{B}$ in Eq.(\ref{V-total}) represents interactions among the bosons,
 which have the same forms as those given in the GWS theory.
 A remark: Discussions to be given below do not depend on explicit expressions of
 the interaction amplitudes in $V$ and, hence, it is unnecessary to
 give discussions for their explicit forms.

\subsection{States of observed particles}

 In this section, we discuss generic descriptions of particle states observed in experiments,
 in particular, their mass states as eigenstates of the total Hamiltonian.
 We use $|0\ra$ to indicate the vacuum state.
 The Fock states mentioned in the previous section are described by
 multiplications of creation operators on the vacuum state $|0\ra$.
 For the sake of clearness in presentation,
 we use $|F_{\bp}^{\alpha}\ra$ to indicate the fermionic state $b^{\alpha \dag}_{F \bp}|0\ra$,
 $|\kappa_{\bp}^{+\alpha}\ra$ for $d^{\alpha \dag}_{\kappa \bp}|0\ra$,
 $|\ov\nu_{\bp}^{\alpha}\ra$ for $d^{\alpha \dag}_{\nu \bp}|0\ra$,
 and $|B_{\bp}^{\lambda}\ra$ for the bosonic state $a^{\lambda \dag}_{B \bp}|0\ra$.

 States of particles observed in experiments should possess certain stability,
 such that they may exist during a finite time period.
 Due to the interactions depicted by $V$ in Eq.(\ref{V-total}),
 such states should correspond to certain superpositions of the Fock states.
 In particular, the mass state of an observed particle, with a measured mass denoted by $m_{F(B)}$,
 should be associated with an eigenstate of the total Hamiltonian $H$
 within certain subspace of the total state space.

 As an example, let us consider the mass state of a positron, denoted by $|\Psi_{e^+}^\alpha\ra$,
 satisfying the eigenequation $H |\Psi_{e^+}^\alpha\ra = m_e |\Psi_{e^+}^\alpha\ra$.
 The subspace spanned by Fock states, which includes this state,
 should contain the free particle state $|e_{\bf 0}^{+\alpha}\ra$ with a zero momentum
 and be invariant under the action of the interaction $V$.
 We use $\E^{e^+}$ to indicate the smallest subspace that possesses this property.
 Specifically, $\E^{e^+}$ is spanned by $|e_{\bf 0}^{+\alpha}\ra$ and all those Fock states,
 denoted by $|\psi_\zeta\ra$, for which
\begin{gather}\label{H-connect-e+}
 \la \psi_\zeta|V^m| \zeta_0  \ra \ne 0
\end{gather}
 for some positive integer $m$.
 Due to the $\delta$-functions in the interaction $V$,
 Eq.(\ref{H-connect-e+}) implies that each Fock state in the space $\E^{e^+}$ should have a zero total momentum.
 The state $|\Psi_{e^+}^\alpha\ra$ has the following expansion in Fock states,
\begin{gather}\notag
 |\Psi_{e^+}^\alpha\ra =  \N_{e^+} \Big(  |e_{\bf 0}^{+\alpha}\ra +
 \int d\ww q \sum_{\beta \lambda} c^{eA}_{\beta \lambda}(\bq) |e_{\bq}^{+\beta}\ra |A_{-\bq}^{\lambda}\ra
 \\ + \int d\ww q \sum_{\beta \lambda} c^{eZ}_{\beta \lambda}(\bq) |e_{\bq}^{+\beta}\ra |Z_{-\bq}^{\lambda}\ra
 \notag
 \\ + \int d\ww q \sum_{\beta \lambda} c^{eW}_{\beta \lambda}(\bq) |\ov\nu_{\bq}^{\beta}\ra |W_{-\bq}^{+\lambda}\ra
 + \cdots \Big) \label{Psi-e+-0}
\end{gather}
 where $\N_{e^+}$ is the normalization coefficient.

 The procedure of renormalization imposes no restriction to the values of
 renormalized masses $\epsilon_{F(B) {\bf 0}}$ of free particles.
 Since the experimentally measured masses $m_{F(B)}$ should be independent of the values of $\epsilon_{F(B) {\bf 0}}$,
 there should exist certain relations between the interaction amplitudes and the values of $\epsilon_{F(B) {\bf 0}}$.
 For those particles that are not the focus of this paper, namely, for
 electron, muon, tauon, $Z^0$-boson, $W^\pm$-bosons, and photon,
 we choose the values of $\epsilon_{F(B) {\bf 0}}$ as close as
 possible to their measured values $m_{F(B)}$ and assume that
\begin{gather}\label{}
 \epsilon_{F(B) {\bf 0}} \approx m_{F(B)}
\end{gather}
 for these particles.
 This closeness of $\epsilon_{F(B) {\bf 0}}$ to $m_{F(B)}$
 implies relative weakness of the corresponding interaction amplitudes.
 Then, noting the zero mass of photon and the largeness of $m_{W}$ and $m_{Z}$,
 it is reasonable to assume that, taking a major part of the right hand side (rhs) of Eq.(\ref{Psi-e+-0}),
 the state $|\Psi_{e^+}^\alpha\ra$ can be written in the following form,
\begin{gather}
  |\Psi_{e^+}^\alpha\ra \approx \N_{e^+}\left( |e_{\bf 0}^{+\alpha}\ra +
 \int d\ww q \sum_{\beta \sigma} c^{e}_{\beta \sigma}(\bq) |e_{\bq}^{+\beta}\ra |f^\sigma_{e^+}\ra
 \right) ,  \label{Psi-e+}
\end{gather}
 where $c^{e}_{\beta \sigma}(\bq) $ are coefficients and
 $|f^\sigma_{e^+}\ra$ represent states of photons and of $(\kappa,\kappa^+)$ pairs,
 with a label $\sigma$ standing for the particle species in $|f^\sigma_{e^+}\ra$
 and their momentum and spin properties.

 Following arguments similar to those given above, the mass state of an antimuon is
 approximately written as
\begin{gather}
 |\Psi_{\mu^+}^\alpha\ra \approx \N_{\mu^+} \left( |\mu_{\bf 0}^{+\alpha}\ra +
 \int d\ww q \sum_{\beta \sigma} c^{\mu }_{\beta \sigma}(\bq) |\mu_{\bq}^{+\beta}\ra |f^\sigma_{\mu^+}\ra
 \right).  \label{Psi-mu+}
\end{gather}
 And,  the mass state of a $W^+$-boson is written as
\begin{gather}
  |\Psi_{W^+}^\alpha\ra \approx \N_{W^+} \left( |W_{\bf 0}^{+\alpha}\ra +
 \int d\ww q \sum_{\beta \sigma} c^{W }_{\beta \sigma}(\bq)
 |W_{\bq}^{+\beta}\ra |f^\sigma_{W^+}\ra \right).
 \label{Psi-W+}
\end{gather}
 By Lorentz transformations, the above states can be transformed to ones
 that have a total momentum $\bp$, which we write as
 $|\Psi_{e^+}^\alpha(\bp)\ra, |\Psi_{\mu^+}^\alpha(\bp)\ra$,
 and $|\Psi_{W^+}^\lambda(\bp)\ra$, respectively.
 The related subspaces are written as, say, $\E^{e^+}(\bp)$.

 Finally, we discuss the focus of this paper, namely, neutrino states.
 Similar to the case of positron discussed above,
 a neutrino mass state is associated with an eigenstate of $H$ within a subspace of the total state space.
 This subspace, denoted by $\E^{\nu}$, includes a basic neutrino state $b_{\nu {\bf 0}}^{\alpha \dag}|0\ra$
 with a zero momentum and is invariant under the interaction $V$.
 For brevity, we use $|\zeta_0\ra$ to indicate this state $b_{\nu {\bf 0}}^{\alpha \dag}|0\ra$,
\begin{gather} \label{zeta-0}
 |\zeta_0\ra := b_{\nu {\bf 0}}^{\alpha \dag}|0\ra, \quad
 H_0|\zeta_0\ra = \epsilon_{\nu {\bf 0}} |\zeta_0\ra.
\end{gather}
 The subspace $\E^{\nu}$ is spanned by the state $|\zeta_0\ra$ and all those Fock states
 $|\psi_\zeta\ra$ that can be ``connected'' to $|\zeta_0\ra$
 through the interaction $V$, such that
\begin{gather}\label{H-connect}
 \la \psi_\zeta|V^m| \zeta_0  \ra \ne 0
\end{gather}
 for some positive integer $m$.
 Clearly, the space $\E^{\nu}$ is invariant under the action of the Hamiltonian $H$,
 i.e., $H|\Psi\ra \in \E^{\nu}$ for all vectors $|\Psi\ra \in \E^{\nu}$.

 A main purpose of this paper is to study the parts of neutrino masses,
 which come from interactions between renormalized states of the basic neutrino and  other particles.
 For this reason, we choose the renormalized mass $\epsilon_{\nu {\bf 0}}$ of free basic neutrino
 as small as possible.

 The space $\E^\nu$ is infinitely large and
 it is a very difficult task to completely solve the eigenproblem of $H$ within it.
 Here,  we restrict ourself to approximate expressions
 for a few lowest-lying eigenstates of $H$ only,
 denoted by $|\Psi_{s}\ra$ with a label $s=1,2,3,\ldots$,
\begin{gather}\label{Hr-ES-m}
 H |\Psi_{s}\ra = m_{ s} |\Psi_{s}\ra, \quad |\Psi_s \ra \in \E^\nu.
\end{gather}
 If such states can be found, as discussed previously,
 we interpret them as mass states of neutrino with masses $m_{s}$.
 As is well known, particle masses are independent of their spin states.
 For this reason, for brevity, we do not explicitly indicate spin states of the vectors $|\Psi_s\ra$.

 For the model proposed above to be of practical interest, it should be able to predict some states,
 which may be interpreted as flavor neutrino states.
 In the next section, we discuss such flavor states.
 Then, in the section after the next, we discuss approximate expressions of the mass states $|\Psi_{s}\ra$.

\section{Flavor neutrino states}\label{sect-e-neutrino}

 In this section, we discuss states that may be related to flavor neutrino states.
 Specifically, we discuss relevant subspaces in  Sec.\ref{sect-subspace-e-neu},
 then, discuss ground states of $H$ within these subspace in Sec.\ref{prop-e-neu},
 and finally in Sec.\ref{sect-F-neutrino} we show that these states have properties
 that are usually expected for flavor neutrino states.

\subsection{Subspaces of $\E^\nu$ related to charged leptons $\kappa$}\label{sect-subspace-e-neu}

 We consider subspaces of $\E^\nu$, denoted by $\E^\nu_{Z\kappa}$ with $\kappa=e, \mu, \tau$,
 which are related to the following parts of the total Hamiltonian $H$,
\begin{gather}\label{H-nu-e}
 H^{\nu}_{Z\kappa} = H_0+ V_{Z} + V_{\kappa}.
\end{gather}
 It is sometimes convenient to give the common part of $H^{\nu}_{Z\kappa}$
 an explicit notation, which we write 
\begin{gather}\label{}
 H^\nu_{Z} = H_0+ V_{Z}.
\end{gather}
 Specifically, the subspace $\E^\nu_{Z\kappa}$ is spanned by $|\zeta_0\ra$
 and all those Fock states $|\psi_\zeta\ra$, for each of which
\begin{gather}\label{VZ-connect}
 \la \psi_\zeta|(V_{Z}+ V_\kappa)^m| \zeta_0  \ra \ne 0
\end{gather}
 for some positive integer $m$.
 By a Lorentz transformation, which changes a zero momentum
 to a momentum $\bp$, the subspace $\E^\nu_{Z\kappa}$ is changed to a
 subspace whose states have a total momentum $\bp$;
 we use $\E^\nu_{Z\kappa}(\bp)$ to denote the new subspace.

 From the forms of the interaction terms $V_Z$ and $V_{\kappa}$
 given previously, it is not difficult to see that,
 in addition to the one-particle state $|\zeta_0\ra$,
 the space $\E^\nu_{Z\kappa}$ is spanned by
 the following two-particle states,
\begin{gather}
  |\Phi_i\ra   = |\nu_\bp^\alpha\ra |Z_{-\bp}^\lambda\ra, \quad
 |\Phi_k\ra  = |\kappa_{\bp}^\alpha \ra |W_{-\bp}^{+\lambda}\ra, \label{2-part-phi}
\end{gather}
 with subscripts $i$ and $k$ standing for $(\bp, \alpha, \lambda)$,
 and states with more than two particles.
 One can write
\begin{gather}\label{S-nu-e}
 \E^\nu_{Z\kappa} = \E^\nu_{Z} \bigoplus \E^\nu_\kappa,
\end{gather}
 where $\E^\nu_\kappa$ is spanned all those Fock states $|\psi_\zeta\ra$ in $\E^\nu_{Z\kappa}$
 that contain the charged lepton $\kappa$, and $\E^\nu_Z$ is spanned
 by the rest Fock states in $\E^\nu_{Z\kappa}$.
 Clearly, the space $\E^\nu_{Z}$ contains $|\zeta_0\ra$ and $|\Phi_i\ra$,
 and the space $\E^\nu_{\kappa}$ contains $|\Phi_k\ra$.


 We assume that $H^{\nu}_{Z\kappa}$ has eigenstates with positive eigenvalues within the space $\E^\nu_{Z\kappa}$,
 and use $|\nu_\kappa\ra$ to indicate the eigenstate with the lowest positive eigenvalue
 denoted by $E_{\nu_\kappa}^0$,
\begin{gather}\label{HnuZk-eigen}
 H^{\nu}_{Z\kappa} |\nu_\kappa\ra  = E_{\nu_\kappa}^0|\nu_\kappa\ra.
\end{gather}
 By Lorentz transformations, these states $|\nu_\kappa\ra$ can be transformed to ones
 that have a total momentum $\bp$, which we indicate by $|\nu_\kappa(\bp)\ra$.
 For the simplicity in discussion, we do not discuss possible degeneracy of the
 eigenvalues $E_{\nu_\kappa}^0$, which can be dealt with by introducing an
 additional quantum number such as spin.

\subsection{Properties of $|\nu_\kappa\ra$ and $ E_{\nu_\kappa}^0$}\label{prop-e-neu}

 In this section, we discuss properties of the three states $|\nu_\kappa\ra$.
 We assume that, with $H_0$ taken as the unperturbed Hamiltonian
 and with $(V_{Z} + V_{\kappa})$ as the perturbation,
 the state $|\nu_\kappa\ra$ has a convergent perturbation expansion,
\begin{gather}\label{nu-k-pert-expan}
 |\nu_\kappa\ra = \N_\kappa \Big( |\nu_\kappa^{(0)}\ra + |\nu_\kappa^{(1)}\ra + \cdots \Big),
\end{gather}
 where $\N_\kappa$ is the normalization coefficient.
 The zeroth-order term is given by $|\zeta_0\ra$, i.e.,  $|\nu_\kappa^{(0)}\ra = |\zeta_0\ra$.

 The first-order contribution $|\nu_\kappa^{(1)}\ra$ is written as
\begin{gather}\label{|nu-e-1>}
  |\nu_\kappa^{(1)}\ra  = \sum_{i} \frac{ H_{  i   0}}{\epsilon_{\nu {\bf 0}}-H_{ii}} |\Phi_i\ra
  + \sum_{k} \frac{ H_{k 0}}{\epsilon_{\nu {\bf 0}}-H_{kk}} |\Phi_k\ra,
\end{gather}
 where $|\Phi_i\ra$ and $|\Phi_k\ra$ are defined in Eq.(\ref{2-part-phi}),
 $H_{i(k)0} = \la\Phi_{i(k)}|H^{\nu}_{Z\kappa}|\zeta_0\ra$,
 $H_{ii} = \la\Phi_i|H_0|\Phi_i\ra = \epsilon_{\nu \bp} + \epsilon_{Z -\bp}$,
 and $H_{kk}= \la\Phi_k|H_0|\Phi_k\ra = \epsilon_{\kappa \bp} + \epsilon_{W^+ -\bp}$.
 Since the masses $m_W$ and $m_Z$ of the bosons $W^\pm$ and $Z^0$
 are very large and the value of $\epsilon_{\nu {\bf 0}}$ is small as discussed previously,
 we assume that the main contribution to the rhs of Eq.(\ref{|nu-e-1>})
 comes from momentum not high, such that $H_{ii}$ and $H_{kk}$ can be approximated by
 $m_Z$ and $m_W + m_\kappa$, respectively.
 Then, $|\nu_\kappa^{(1)}\ra$ can be approximately written in the following form,
\begin{gather}\label{nu-e1}
 |\nu_\kappa^{(1)}\ra \simeq \frac{1}{m_Z}|\varphi_{\kappa 1}\ra
 + \frac{1}{m_W + m_\kappa}|\varphi_{\kappa 2}\ra ,
\end{gather}
 where $|\varphi_{\kappa 1}\ra \in \E^\nu_{Z}$ comes from the first summation on
 the rhs of Eq.(\ref{|nu-e-1>}) and $|\varphi_{\kappa2}\ra \in \E^\nu_\kappa$
 comes from the second summation.
 Below, we neglect contributions from the second- and higher-order
 perturbation-expansion terms of $|\nu_\kappa\ra$.

 We find that the two terms on the rhs of Eq.(\ref{nu-e1}) should have close norms,
 as discussed below.
 The form of the bare Lagrangian in Eq.(\ref{Lag}) implies that
 the effective interaction Hamiltonian for the interaction between $|\zeta_0\ra$ and
 $|\Phi_i\ra  \in \E^\nu_Z$ should have the following form,
\begin{gather}\label{}
 \HH_{1} = - \frac{\ww g}{2 \cos \theta_{\rm w}}   \ww{\ov\nu}_L\gamma^\mu \ww\nu_L \ww Z^0_\mu ,
\end{gather}
 where $\ww\nu_L$ and $\ww Z^0_\mu$ represents the renormalized
 neutrino field and the renormalized $Z^0$-boson field, respectively,
 $\theta_{\rm w}$ is the Weinberg angle, and $\ww g$ is a renormalized coupling constant.
 Meanwhile, the effective interaction Hamiltonian for the coupling between $|\zeta_0\ra$
 and $|\Phi_k\ra  \in \E^\nu_\kappa$ is written as
\begin{gather}\label{}
 \HH_{2} = - \frac {\ww g}{\sqrt 2 } \Big( \ww{ \ov \nu}_L \gamma^\mu \ww \kappa_L \ww W^+_\mu
  + \ww{\ov \kappa}_L \gamma^\mu \ww\nu_L \ww W^-_\mu \Big),
\end{gather}
 where $\ww\kappa_L$ and $\ww W^\pm_\mu$ represent the
 renormalized lepton $\kappa$ field and the renormalized $W^\pm$-boson field, respectively.

 From the forms of $\HH_{1}$ and $\HH_{2}$ given above, it is seen that
 prefactors of the elements $H_{i0}$ and $H_{k0}$ should have a difference of $1/(\sqrt 2 \cos\theta_{\rm w})$.
 Noting the relation $m_W = m_Z \cos\theta_{\rm w}$,
 the difference between the two factors $1/m_Z$ and $1/(m_W+m_\kappa) \simeq 1/m_W$ on the rhs of Eq.(\ref{nu-e1})
 compensates the $\cos\theta_{\rm w}$ part in the factor $1/(\sqrt 2 \cos\theta_{\rm w})$.
 Moreover, in the above expressions the left-handed neutrino field is written in the same way as in the
 SM, where it has no right-handed part; this gives rise to a factor $1/\sqrt 2$
 in its normalization coefficient compared with that of the $\kappa$ field.
 This factor $1/\sqrt 2$ compensates the $1/\sqrt 2$ part in $1/(\sqrt 2 \cos\theta_{\rm W})$.
 Therefore, the two terms on the rhs of Eq.(\ref{nu-e1}),
 i.e., $\frac{1}{m_Z}|\varphi_{\kappa 1}\ra$ and $\frac{1}{m_W + m_\kappa}|\varphi_{\kappa 2}\ra$
 should have similar expansion coefficients
 in the basis $|\Phi_i\ra$ and $|\Phi_k\ra$, respectively.
 This implies that
\begin{gather}\label{varphi-norms}
  \frac{1}{m_Z^2} \la \varphi_{\kappa 1}|\varphi_{\kappa 1}\ra
 \simeq \frac{1}{m_W^2}  \la \varphi_{\kappa 2}| \varphi_{\kappa 2}\ra.
\end{gather}


 For the sake of convenience in later discussions, we write the state $|\nu_\kappa\ra$ as
\begin{gather}\label{nu-kappa}
 |\nu_\kappa\ra = |\varphi_0\ra + |\eta_\kappa\ra,
\end{gather}
 where $|\varphi_0\ra  \in \E^\nu_{Z}$ and $|\eta_\kappa\ra \in \E^\nu_\kappa$.
 Up to the first-order perturbation expansion, the above discussions give that
\begin{subequations}\label{varphi0-eta-ka}
\begin{gather}\label{}
 |\varphi_0\ra \simeq \N_\kappa  \left(|\zeta_0\ra + \frac{1}{m_Z}|\varphi_{\kappa 1}\ra \right),
 \\ |\eta_\kappa\ra \simeq  \frac{\N_\kappa}{m_W + m_\kappa}|\varphi_{\kappa 2}\ra.
\end{gather}
\end{subequations}
 If the state $|\zeta_0\ra$ has a small population in the state $|\nu_\kappa\ra$,
 then, Eq.(\ref{varphi-norms}) implies that,
\begin{gather}\label{varphi0-eta-kappa}
 \la\varphi_0|\varphi_0\ra \simeq \la \eta_\kappa|\eta_\kappa\ra.
\end{gather}

 Now, let us compute the eigenvalue $E_{\nu_\kappa}^0$,
 given by $E_{\nu_\kappa}^0 = \la \nu_\kappa|H^{\nu}_{Z\kappa}|\nu_\kappa\ra$.
 Simple derivation shows that
\begin{gather}\label{m-kappa-nu-W}
 E_{\nu_\kappa}^0  \simeq \epsilon_{\nu {\bf 0}}' + \frac{G_{1}}{m_Z} +  \frac{G_{2}}{m_W + m_\kappa},
\end{gather}
 where $\epsilon_{\nu {\bf 0}}' = \N_\kappa^2 \epsilon_{\nu {\bf 0}}$ and
\begin{gather}\label{G1}
 G_{1} = 2 \N_\kappa^{2} \Re \Big(\la\zeta_0|V_Z|\varphi_{\kappa 1}\ra \Big)
 +\N_\kappa^2 \frac{\la\varphi_{\kappa 1}|H^\nu_{Z}|\varphi_{\kappa 1}\ra}{m_Z},
 \\ G_{2} = 2\N_\kappa^2\Re \left(\la\zeta_0|V_\kappa|\varphi_{\kappa 2}\ra
 + \frac{\la\varphi_{\kappa 1}|H^\nu_{Z\kappa}|\varphi_{\kappa 2}\ra}{m_Z} \right) \notag
 \\ + \N_\kappa^2\frac{\la\varphi_{\kappa 2}|H^\nu_{Z\kappa}|\varphi_{\kappa 2}\ra}{m_W+m_\kappa}, \label{G2}
\end{gather}
 with $\Re( \ )$ indicating the real part of $(\ )$.
 Since the basic neutrino interacts in the same way with the three charged leptons $\kappa$,
 the dependence of $|\varphi_0\ra$, $G_1$, and $G_2$ on $\kappa$ should comes from the masses $m_\kappa$.
 Then, noting that $m_W \gg m_\kappa$, from
 previous discussions about the vectors $|\varphi_{\kappa 1}\ra$ and $|\varphi_{\kappa 2}\ra$
 and  Eqs.(\ref{G1})-(\ref{G2}), it is seen  that
 the vector $|\varphi_0\ra$ and the two quantities $G_1$ and $G_2$
 should be approximately independent of $\kappa$.
 This implies that
\begin{gather}\label{nu-3m}
 E_{\nu_e}^0 \approx E_{\nu_\mu}^0 \approx E_{\nu_\tau}^0.
\end{gather}
 More exactly, direct derivation shows that
\begin{gather} \label{m-nu-diff}
 E_{\nu_\kappa}^0- E_{\nu_{\kappa'}} \approx \frac{m_{\kappa'} - m_\kappa}{m_W^2} G_2.
\end{gather}

\subsection{$|\nu_\kappa\ra$ as flavor neutrino states}\label{sect-F-neutrino}

 In this section, we discuss conditions, under which the states
 $|\nu_\kappa\ra$ in Eq.(\ref{nu-kappa}) can be associated with flavor neutrino states.
 Below, for brevity, we indicate the three flavor neutrinos as \emph{$\kappa$-neutrinos} with $\kappa =e, \mu$, and $\tau$.

 Let us first discuss some subspaces in which flavor neutrino states may lie.
 To find out such subspaces,  one method is to study those states
 that the interaction $V$ may generate
 from a $W^+$-boson state $|\Psi_{W^+}^\lambda\ra$ in Eq.(\ref{Psi-W+}),
 in the accompany of a charged lepton $\kappa$.
 To be specific, let us discuss, say, a subspace for an $e$-neutrino;
 it is spanned by those states that may be generated in processes of the form $W^+ \to e^+ + \text{$e$-neutrino}$,
 in the accompany of positron states $|\Psi_{e^+}^\alpha(\bp)\ra$.
 To find out possible configurations of the Fock states for the subspace,
 it is unnecessary to write complete expressions of the Fock states
 and, for brevity, we just write, e.g., $|e^+\ra$ for $|e^{+\alpha }_\bp\ra$.
 It is not difficult to check that the process, $W^+ \to e^+ + \text{$e$-neutrino}$,
 generates the following configurations of Fock states,
\begin{gather}\label{confg-fromW+}\notag
 |e^+\ra |\nu\ra, \ |e^+\ra |\nu\ra |A\ra, \ |e^+\ra |\nu\ra |Z\ra, \ |W^+\ra |\ov\nu\ra |\nu\ra,
 \\ |e^+\ra |W^+\ra |e\ra, \ |W^+\ra |e^+\ra |e\ra, \ldots .
\end{gather}
 Furthermore, it is not difficult to verify that all these states are contained in direct-product spaces of
 the type $ \E^{e^+}(\bp) \otimes \E^\nu_{Z e}(-\bp)$.

 Therefore, it is reasonable to expect that a flavor $e$-neutrino state
 with a zero momentum should lie in the subspace $\E^\nu_{Z e}$.
 The most natural candidate in this subspace for this flavor neutrino state
 is the state that has the lowest energy expectation value.
 This is just the state $|\nu_e\ra$ discussed in the previous two sections.
 Similarly, states of the flavor $\mu$-neutrino and $\tau$-neutrino with zero total momentum
 should lie in the subspaces $\E^\nu_{Z \mu}$ and $\E^\nu_{Z \tau}$, respectively,
 with $|\nu_\mu\ra$ and $|\nu_\tau\ra$ as the most natural candidates.

 Below, we study whether the states $|\nu_\kappa\ra$ may have the following
 well-known property of flavor neutrinos, that is,  a $\kappa$-neutrino
 interacts effectively with the charged lepton $\kappa$ only, or almost so.
 This is a nontrivial problem in the proposed model,
 because there is only one basic neutrino
 and the three states $|\nu_\kappa\ra$ are not orthogonal to each other.
 In fact, they have an overlap given by $\la\nu_\kappa|\nu_{\kappa'}\ra
 = \la\varphi_0|\varphi_0\ra$ for $\kappa \ne \kappa'$.

 Without loss of generality, let us study, say,
 whether the transition $\nu_e + \mu^+ \to W^+$ may have a very small amplitude,
 in contrast to the transition $\nu_e + e^+ \to W^+$ which may have a notable amplitude.
 For a process $\nu_e + e^+ \to W^+$, the initial states of the the two incoming particles,
 before the interaction happens effectively, can be taken as $|\nu_e(\bp)\ra$
 (for the $e$-neutrino) and $|\Psi_{e^+}^\alpha(\bq)\ra$ (for the positron).
 What is of interest is the probability amplitude for a $W^+$-boson to come out,
 which is in a state $|\Psi_{W^+}^\lambda (\bk)\ra$ with $\bk = \bp +\bq$.
 This probability amplitude corresponds to an element of the S-matrix.
 As is well known, such probability amplitudes can be computed by Feynman's diagrams.

 The amplitude for the basic Feynman diagram of $\nu_e + e^+ \to W^+$,
 denoted by $G_{ W^+}^{\nu_e e^+} $, is written as
\begin{gather}\label{G-e+}
 G_{ W^+}^{\nu_e e^+} = \la \Psi_{W^+}^\lambda (\bk)|H|\nu_e(\bp)\ra |\Psi_{e^+}^\alpha(\bq)\ra.
\end{gather}
 Similarly, the amplitude of the basic Feynman diagram
 of $\nu_e + \mu^+ \to W^+$ is written as
\begin{gather}\label{G-mu+}
 G_{ W^+}^{\nu_e \mu^+} = \la \Psi_{W^+}^\lambda (\bk)|H|\nu_e(\bp)\ra |\Psi_{\mu^+}^\alpha(\bq)\ra.
\end{gather}
 Below, we compare these two interaction amplitudes.

 According to Eq.~(\ref{nu-kappa}), the state $|\nu_e\ra$ can be written as
\begin{gather}\label{nue-'}
  |\nu_e\ra = a|\zeta_0\ra + |\varphi_0'\ra + |\eta_e\ra,
\end{gather}
 where $a = \la \zeta_0|\varphi_0\ra$ and
 $|\varphi_0'\ra = |\varphi_0\ra - a|\zeta_0\ra$.
 Note that $|\varphi_0'\ra$ is composed of vectors $|\Phi_i\ra$ in Eq.~(\ref{2-part-phi}),
 each containing a $Z^0$-boson and a basic neutrino, meanwhile,
 $|\eta_e\ra $ is composed of $|\Phi_k\ra$, each containing a $W^+$-boson and an electron.
 Substituting the expressions in Eqs.(\ref{Psi-e+}), (\ref{Psi-W+}), and (\ref{nue-'}),
 with appropriate change of the total momenta, into Eq.(\ref{G-e+}),
 it is not difficult to find that contributions from terms with $V_Z$ and most terms with $H_0$ and $V_B$ are zero,
 and the same for contributions from $|\varphi_0'\ra$.
 The final result can be  schematically written as
\begin{gather}\label{}\notag
 G_{ W^+}^{\nu_e e^+} \approx  a\la W^+|V_e|\zeta_0\ra |e^+\ra
 \\ + \sum \int a \la f_{W^+}| \la W^+| V_e|\zeta_0\ra|e^+\ra |f_{e^+}\ra \notag
 \\ + \la f_{W^+}| \la W^+| H_0 +V_e +V_B|\eta_e\ra \Big( |e^+\ra + |e^+\ra |f_{e^+}\ra \Big), \label{G-e-app1}
\end{gather}
 where, for brevity, labels for momentum and spin are not written explicitly.

 Following a similar procedure, one finds that the amplitude $G_{ W^+}^{\nu_e \mu^+}$ is
 schematically written as
\begin{gather}\notag
 G_{ W^+}^{\nu_e \mu^+} \approx  a\la W^+|V_\mu|\zeta_0\ra |\mu^+\ra
 \\ + \sum \int a \la f_{W^+}| \la W^+| V_\mu|\zeta_0\ra|\mu^+\ra |f_{\mu^+}\ra. \label{G-mu-app1}
\end{gather}
 From Eqs.(\ref{G-e-app1}) and (\ref{G-mu-app1}),
 it is seen that, if the quantity $a$ is very small,
 then, the interaction amplitude $G_{ W^+}^{\nu_e \mu^+}$ is very small,
 meanwhile, the amplitude $ G_{ W^+}^{\nu_e e^+}$ usually has notable values.

 Following arguments similar to those given above, one finds that, as long as the quantity $a$ remains small,
 all the interaction amplitudes $G_{ W^+}^{\nu_\kappa {\kappa'}^+}$
 with $\kappa \ne \kappa'$ are small.
 In other words, the probabilities of the processes $\nu_\kappa + {\kappa'}^+ \to W^+$ with
 $\kappa \ne \kappa'$ are much suppressed,
 compared with those processes with $\kappa = \kappa'$.
 This suggests that the states $|\nu_\kappa\ra$ may be interpreted as flavor $\kappa$-neutrino states.

\section{Mass states of neutrino}\label{sect-mass-state}

 In this section, making use of the above-obtained expressions of the three states $|\nu_\kappa\ra$,
 we study properties of the neutrino mass states $|\Psi_s\ra$ in the space $\E^\nu$.
 Since the values $E_{\nu_\kappa}^0$ are close to each other [see Eq.(\ref{nu-3m})],
 usually one can not get approximate expressions of the states $|\Psi_s\ra$
 by a first-order perturbation treatment to $|\nu_\kappa\ra$.

 To find approximate expressions for $|\Psi_s\ra$, we employ a variational approach.
 Specifically, in Sec.\ref{sect-delta-F} we compute the variation of a relevant
 functional at a specific state vector;
 then, in Sec.\ref{sect-psi-s}, making use of the result obtained, we find
 approximate expressions for neutrino mass states; finally, in Sec.\ref{sect-mass-s},
 we derive approximate ratios of neutrino mass differences and compare  them
 with experimental results.

\subsection{$\delta F$ for a specific vector}\label{sect-delta-F}

 In a variational approach to eigenstates of a Hamiltonian $H$,
 one computes variation of a functional $F$,
\begin{gather}\label{}
 F = \la\psi|H|\psi\ra - \lambda \la\psi|\psi\ra
\end{gather}
 for an unnormalized vector $|\psi\ra$.
 Below, we compute $\delta F$ around the following vector,
\begin{gather}\label{psi}
 |\psi\ra = |\varphi_0\ra + \sum_\kappa c_\kappa |\eta_\kappa\ra,
\end{gather}
 where $c_\kappa$ are real parameters satisfying
\begin{gather}\label{sum-c=1}
 \sum_{\kappa} c_\kappa =1.
\end{gather}
 In this section, we compute the variation within the following subspace of $\E^{\nu}$,
\begin{gather}\label{}
 \E^\nu_1 = \E^\nu_{Z} \oplus \E^\nu_{e}  \oplus \E^\nu_{\mu}  \oplus \E^\nu_{\tau}.
\end{gather}
 The results to be obtained will be used in the next section when deriving
 an approximate expression for the mass states $|\Psi_s\ra$.

 It proves convenient to first discuss the following functionals,
\begin{gather}\label{}
 F_\kappa = \la \nu_\kappa|H^\nu_{Z\kappa}|\nu_\kappa\ra - E_{\nu_\kappa}^0
 \la \nu_\kappa|\nu_\kappa\ra.
\end{gather}
 Equation (\ref{HnuZk-eigen}) implies that $\delta F_k =0$ under an arbitrary variation in the space $\E^\nu_{Z\kappa}$,
 given by $|\varphi_0\ra \to |\varphi_0\ra + |\delta\varphi_0\ra$
 and $c_\kappa|\eta_\kappa\ra \to c_\kappa|\eta_\kappa\ra + |\delta\eta_\kappa\ra$,
 where $ | \delta \varphi_0\ra \in \E^\nu_Z$ and $|\delta \eta_\kappa\ra \in \E^\nu_\kappa$.
 It is not difficult to verify that
\begin{gather}\label{} \notag
 \delta F_k = 2\Re \Big( \la \delta \varphi_0|H^{\nu}_{Z}|\varphi_0\ra +\la \delta \varphi_0|V_\kappa|\eta_\kappa\ra
 + \la \delta\eta_\kappa|H^{\nu}_{Z}|\eta_\kappa\ra
 \\ + \la \delta\eta_\kappa|V_\kappa|\varphi_0\ra
  + \la \delta\eta_\kappa|V_\kappa|\eta_k\ra \Big) \notag
  \\ - E_{\nu_\kappa}^0 \Re \Big(\la \delta \varphi_0|\varphi_0\ra + \la \delta\eta_\kappa|\eta_\kappa\ra \Big).
 \label{Re-delta-Hk}
\end{gather}
 In the derivation of Eq.(\ref{Re-delta-Hk}), the following properties have been used:
\begin{gather}\label{3prop-used}
  \begin{array}{lll}
    (a) & \hbox{$H^{\nu}_{Z}|\varphi_0\ra \ \& \  H^{\nu}_{Z}|\delta\varphi_0\ra \in \E^\nu_{Z}$;} \\
    (b) & \hbox{the space $\E^\nu_{Z}$ is orthogonal to $\E^\nu_{\kappa}$;} \\
    (c) & \hbox{$\la \delta \varphi_0|V_\kappa|\varphi_0\ra =0$.}
  \end{array}
\end{gather}

 A generic variation of the vector $|\psi\ra$ in the space $\E^{\nu}_1$
 is written as $\delta |\psi\ra = |\delta\varphi_0\ra + \sum_\kappa |\delta\eta_\kappa\ra$.
 This gives $ \delta \la\psi|\psi\ra= 2\Re (\la \delta \varphi_0|\varphi_0\ra + \sum_\kappa c_\kappa
 \la \delta\eta_\kappa|\eta_\kappa\ra )$.
 Noting Eq.(\ref{sum-c=1}), one gets that
\begin{gather}
 \delta \la\psi|\psi\ra= 2\Re \left[ \sum_\kappa c_\kappa \big(\la \delta \varphi_0|\varphi_0\ra +
 \la \delta\eta_\kappa|\eta_\kappa\ra \big) \right] .
\end{gather}
 Let us write
\begin{gather}\label{}\notag
 \la \delta \psi|H|\psi\ra
  = \left(\la\delta \varphi_0| + \sum_{\kappa_1}  \la\delta\eta_{\kappa_1}| \right)
  \left(H^{\nu}_{Z} + \sum_\kappa V_{\kappa}\right)
 \\ \times \left(|\varphi_0\ra + \sum_{\kappa_2} c_{\kappa_2} |\eta_{\kappa_2}\ra\right). \label{<dp|Hp>-1}
\end{gather}
 Noting the properties in (\ref{3prop-used}), the rhs of Eq.(\ref{<dp|Hp>-1}) is further written as
\begin{gather*}
  \la\delta \varphi_0| H^{\nu}_{Z} |\varphi_0\ra
 + \la\delta \varphi_0| \sum_\kappa V_{\kappa}\sum_{\kappa_2} c_{\kappa_2} |\eta_{\kappa_2}\ra
 \\ + \sum_{\kappa_1} \la\delta\eta_{\kappa_1}|
 H^{\nu}_{Z} \sum_{\kappa_2} c_{\kappa_2} |\eta_{\kappa_2}\ra
  +  \sum_{\kappa_1}  \la\delta\eta_{\kappa_1}|
 \sum_\kappa V_{\kappa}|\varphi_0\ra
 \\ + \sum_{\kappa_1}  \la\delta\eta_{\kappa_1}|
 \sum_\kappa V_{\kappa} \sum_{\kappa_2} c_{\kappa_2} |\eta_{\kappa_2}\ra.
\end{gather*}
 Making use of the fact that none of $\la\delta \varphi_0|$, $|\varphi_0\ra$, and $H^{\nu}_{Z}$
 contains any particle $\kappa$, it is easy to see that
\begin{gather}
 \la\delta \varphi_0| \sum_\kappa V_{\kappa} \sum_{\kappa_2} c_{\kappa_2} |\eta_{\kappa_2}\ra
 = \sum_\kappa c_{\kappa}  \la\delta \varphi_0|  V_{\kappa} |\eta_{\kappa}\ra,
 \\ \sum_{\kappa_1}  \la\delta\eta_{\kappa_1}|
 H^{\nu}_{Z}  \sum_{\kappa_2} c_{\kappa_2} |\eta_{\kappa_2}\ra
 = \sum_\kappa c_{\kappa} \la\delta\eta_{\kappa}| H^{\nu}_{Z} |\eta_{\kappa}\ra,
 \\   \sum_{\kappa_1}  \la\delta\eta_{\kappa_1}|
  \sum_\kappa V_{\kappa} |\varphi_0\ra
 =  \sum_{\kappa}  \la\delta\eta_{\kappa}|   V_{\kappa} |\varphi_0\ra.
\end{gather}
 Furthermore, $\la\delta\eta_{\kappa_1}|V_\kappa|\eta_{\kappa_2}\ra =0$
 if $\kappa \ne \kappa_1$ or $\kappa \ne \kappa_2$, as a result,
\begin{gather}
  \sum_{\kappa_1}  \la\delta\eta_{\kappa_1}|
 \sum_\kappa V_{\kappa} \sum_{\kappa_2} c_{\kappa_2} |\eta_{\kappa_2}\ra
  =  \sum_\kappa c_{\kappa} \la\delta\eta_{\kappa}|  V_{\kappa} |\eta_{\kappa}\ra.
\end{gather}

 Making use of the results given above, direct derivation shows that
\begin{gather}\label{delta-F}
 \delta F = \sum_\kappa c_\kappa \delta F_\kappa + \delta f_1 + \delta f_2,
\end{gather}
 where
\begin{gather}\label{delta-f1}
 \delta f_1 = \sum_\kappa 2\Re \Big[ c_\kappa (E_{\nu_\kappa}^0 -\lambda ) \big(\la \delta \varphi_0|\varphi_0\ra +
  \la \delta\eta_\kappa|\eta_\kappa\ra\big) \Big],
 \\ \delta f_2 = \sum_{\kappa} 2\Re \Big[ (1-c_\kappa) \la\delta\eta_{\kappa}|  V_{\kappa} |\varphi_0\ra
  \Big]. \label{delta-f2}
\end{gather}
 It is not difficult to see that $\delta F_\kappa =0$ under an arbitrary variation in the space $\E^{\nu}_1$,
 hence, $\delta F = \delta f_1 + \delta f_2$.

\subsection{Approximate expressions for neutrino mass states}\label{sect-psi-s}

 In this section, we discuss approximate expressions for neutrino mass states,
 which can be obtained from $|\psi\ra$ in Eq.(\ref{psi}).
 A sufficient and necessary condition for $|\psi\ra$ to be an eigenstate of $H$
 is that $\delta F=0$ under an arbitrary variation in the space $\E^\nu$.

 Besides the above-discussed subspace $\E_1^\nu$, the space $\E^\nu$ contains other subspaces,
 specifically, $\E_2^\nu = \E^\nu_{e}  \otimes \E^\nu_{\mu}$,
 $\E_3^\nu = \E^\nu_{e}  \otimes \E^\nu_{\tau}$,
 $\E_4^\nu = \E^\nu_{\mu}  \otimes \E^\nu_{\tau}$,
 and $\E_5^\nu = \E^\nu_{e}  \otimes \E^\nu_{\mu}  \otimes \E^\nu_{\tau}$.
 We note that, with respect to each species $\kappa$,
 the total Hamiltonian $H$ can be written as
\begin{gather}\label{}
 H = H^\nu_{Z\kappa} + \sum_{\kappa' \ne \kappa} V_{\kappa'},
\end{gather}
 and the subspace $\E^\nu_1$ contains all the ground states of $H^\nu_{Z\kappa}$
 of $\kappa =e,\mu,\tau$.
 Because of these properties, we assume that major parts of a few 
 lowest-lying eigenstates of $H$ in the space $\E^\nu$ may
 lie in this subspace $\E^\nu_1$.
 To find approximate expressions for these eigenstates, one may consider $\delta F$
 within the subspace $\E^\nu_1$,
 which has been computed and given in Eq.(\ref{delta-F}).

 If an eigenvalue $\lambda$ is close to the three $E_{\nu_\kappa}^0$,
 the values of which are close to each other [see Eq.(\ref{nu-3m})],
 then, $\delta f_1$ in Eq.(\ref{delta-f1}) is small.
 Later, we'll see that this assumption about $\lambda$ is self-consistent, in the sense that their values
 to be derived are indeed close to $E_{\nu_\kappa}^0$.
 Therefore, below, we focus on the term $\delta f_2$.

 One can not considerably reduce all the three terms of $\delta f_2$
 on the rhs of Eq.(\ref{delta-f2}) at the same time.
 Let us consider the following sets of $(c_e,c_\mu,c_\tau)$, which we denote
 by $\{c_\kappa^{(s)}\}$ with $s=1,2,3$,
\begin{gather}\label{c(s)}
 (c_e^{(s)}, c_\mu^{(s)}, c_\tau^{(s)}) = \left\{
                                            \begin{array}{ll}
                                             (-1,1,1) & \hbox{for s=1,} \\
                                              (1,-1,1) & \hbox{for s=2,} \\
                                              (1,1,-1) & \hbox{for s=3.}
                                            \end{array}                                          \right.
\end{gather}
 We use $|\psi_s\ra$  to indicate  the vectors $|\psi\ra$ given by the above sets of $\{c_\kappa^{(s)}\}$, namely,
\begin{gather}\label{psi-s}
 |\psi_s\ra = |\varphi_0\ra + \sum_\kappa c_\kappa^{(s)} |\eta_\kappa\ra.
\end{gather}
 Noting Eq.(\ref{varphi0-eta-kappa}), it is easy to
 see that $|\psi_s\ra$ are approximately orthogonal to each other.

 Let us consider one of the three vectors $|\psi_s\ra$ with a fixed value of $s$.
 For the specific values of $c_\kappa^{(s)}$ given in Eq.(\ref{c(s)}),
 $\delta f_2$ vanishes under whatever variation
 within two of the three subspaces of $\E^\nu_\kappa$ for which $c_\kappa^{(s)}=1$.
 In the rest subspace $\E^\nu_\kappa$ with $c_\kappa^{(s)}=-1$,
 one may have notable $\delta f_2$;
 in principle, it should be possible to modify the vector $|\eta_\kappa\ra$
 for this $\kappa$, such that $\delta f_2$ can be considerably reduced (see the appendix).
 However, this improvement may be not so important, because as discussed previously we have neglected variations
 in the subspaces $\E^\nu_i$ of $i=2,3,4,5$.

 Based on discussions given above, it should be reasonable to expect that the states $|\psi_s\ra$
 be close to eigenstates of $H$.
 Since $|\psi_s\ra$ are composed of the lowest-energy states within the subspaces $\E^\nu_{Z\kappa}$,
 we argue that they may be close to mass states of neutrino, namely, $|\psi_s\ra \approx |\Psi_s\ra$, such that
\begin{gather}\label{H-psis-ms}
 H|\psi_s\ra \approx m_s|\psi_s\ra, \quad s=1,2,3.
\end{gather}

\subsection{Neutrino-mass differences}\label{sect-mass-s}

 Now, we discuss properties of neutrino masses.
 Applying the projection operator for the subspace $\E^\nu_Z$, denoted by  $P^\nu_Z$,
 to both sides of Eq.(\ref{H-psis-ms}),
 one gets that
\begin{gather}\label{part-H|varphi0>}
 P^\nu_Z \left[  H^\nu_Z|\varphi_0\ra
 + \sum_\kappa  c^{(s)}_\kappa V_\kappa|\eta_\kappa\ra \right] \approx m_s|\varphi_0\ra,
\end{gather}
 where we have used the facts that
\begin{subequations}\label{PnuZ=0}
\begin{gather}
 P^\nu_Z H^\nu_{Z}|\eta_\kappa\ra =0 ,
 \\ P^\nu_Z V_\kappa|\varphi_0\ra =0,
 \\ P^\nu_Z V_\kappa |\eta_{\kappa'}\ra =0, \ \text{for} \ \kappa \ne \kappa'.
\end{gather}
\end{subequations}
 Making use of Eqs.(\ref{H-nu-e}), (\ref{nu-kappa}), and (\ref{PnuZ=0}),
 it is not difficult to find that
\begin{gather}\label{PZ-Hk=}
  P^\nu_Z H^\nu_{Z\kappa}|\nu_\kappa\ra = P^\nu_Z \big( H^\nu_Z|\varphi_0\ra +  V_\kappa|\eta_\kappa\ra\big).
\end{gather}
 Then, inserting $(\sum_\kappa c_\kappa^{(s)})=1$ before $|\varphi_0\ra$
 on the left-hand side of Eq.(\ref{part-H|varphi0>}) and
 noting Eqs.(\ref{HnuZk-eigen}) and (\ref{PZ-Hk=}), simple derivation gives that
\begin{gather}\label{ms-mk}
m_s \approx \sum_\kappa c_\kappa^{(s)} E_{\nu_\kappa}^0.
\end{gather}
 It is seen that the values of $\lambda$ given by these $m_s$ are indeed close
 to $E_{\nu_\kappa}^0$, which we assumed in the previous section.

 Using the values of $c_\kappa^{(s)}$ given in Eq.(\ref{c(s)}), Eq.(\ref{ms-mk}) gives that
\begin{subequations}\label{m-123-diff}
\begin{gather}
 m_{1}- m_{2} \approx 2E_{\nu_mu}^0 -2E_{\nu_e}^0, \label{m-123-diff-1}
 \\ m_{1}- m_{3} \approx 2E_{\nu_\tau}^0 -2E_{\nu_e}^0,
 \\ m_{2}- m_{3} \approx 2E_{\nu_\tau}^0 -2E_{\nu_mu}^0.
\end{gather}
\end{subequations}
 Substituting Eq.(\ref{m-nu-diff}) into Eq.(\ref{m-123-diff})
 and noting the experimental results that $m_W \gg m_\tau \gg m_\mu \gg m_e$, one finds that
\begin{subequations}\label{m-123-diff-e}
\begin{gather}\label{}
 m_{1}- m_{2} \approx  -2 m_\mu  G_2/m_W^2, \label{m-123-diff-1e}
 \\ m_{1}- m_{3} \approx -2m_\tau G_2/m_W^2,
 \\ m_{2}- m_{3} \approx -2m_\tau G_2/m_W^2.
\end{gather}
\end{subequations}
 Then,  noting Eqs.(\ref{ms-mk}) and (\ref{nu-3m}), one gets that
\begin{gather}\label{m23/m12-theor}
 \frac{\Delta m^2_{23}}{\Delta m^2_{12}} \approx \frac{m_\tau}{m_\mu } \simeq 17,
\end{gather}
 where $\Delta m^2_{ij} = m_i^2 -m_j^2 $ and the experimental results of $m_\mu \simeq 106 {\rm Mev}$ and
 $m_\tau \simeq 1.8 {\rm Gev}$ have been used.

 Global fits to neutrino oscillation measurements give that
  $|\Delta m^2_{12}|_{\rm ex} \simeq 7.6 \times 10^{-5} {\rm eV}^2 $ and
 $|\Delta m^2_{23}|_{\rm ex} \simeq 2.5 \times 10^{-3} {\rm eV}^2$
 \cite{Caldwell17,Capozzi17,Esteban17,Forero14,Gonzalez14}, as a result,
\begin{gather}\label{m23/m12}
  \frac{|\Delta m^2_{23}|_{\rm ex}}{|\Delta m^2_{12}|_{\rm ex}} \simeq 33.
\end{gather}
 Thus, the above theoretical estimate to the ratio of neutrino-mass difference is about half of the
 experimental result.

 Finally, we discuss a possibility of giving a rough estimate to
 the scale of neutrino mass,
 by making use of the above-discussed experimental results of $|\Delta m^2_{12}|_{\rm ex}$
 and $|\Delta m^2_{23}|_{\rm ex}$.
 Note that, according to Eqs.(\ref{nu-3m}) and (\ref{m-123-diff}), the values of $E_{\nu_\kappa}^0$
 and $m_s$ are close to each other.
 Hence, in the discussion of a rough estimate to them, we simply write them as  $m_\nu$.

 If one neglects the term $\epsilon_{\nu 0}'$ in Eq.(\ref{m-kappa-nu-W}), then, one gets that
\begin{gather}\label{G2-app}
 G_2   \approx {\eta m_{\nu}  m_W},
\end{gather}
 where $\eta = G_2/(G_1+G_2)$.
 Substituting Eq.(\ref{G2-app}) into the first and third subequations of Eq.(\ref{m-123-diff-e}),
 one gets the following estimates to $m_\nu$,
\begin{subequations}\label{mi-esti}
\begin{gather}
 m_{\nu} \approx \sqrt{\frac{|\Delta m^2_{12}|_{\rm ex} m_W}{4\eta \ m_\mu}}
 \approx  0.17  {\rm eV} \ \text{for} \ \eta=0.5,
 \\ m_{\nu} \approx \sqrt{\frac{|\Delta m^2_{23}|_{\rm ex} m_W}{4\eta m_\tau}}
 \approx  0.23  {\rm eV} \ \text{for} \ \eta=0.5,
\end{gather}
\end{subequations}
 where the experimental result of $m_W \simeq 81 {\rm Gev}$ has been used.

 The estimates to neutrino mass given in Eq.(\ref{mi-esti}) are in consistence with
 a recent estimate of $\Sigma = m_1 + m_2 + m_3$ obtained from cosmological data,
 which suggests that $0.05{\rm eV} \lesssim \Sigma \lesssim 1 {\rm eV}$ \cite{arX17-Capo}.
 They lie in the same order of magnitude as results of some other analyses of cosmological data,
 which suggest upper bounds of $\Sigma$ as
 $0.18 {\rm eV}$ \cite{Sunny16} and $0.15 {\rm eV}$ \cite{Sunny17}.

\section{Summary and discussions}\label{sect-conclusion}

\subsection{Summary}

 In this paper, a model is proposed, in which there is only one basic neutrino
 at the fundamental level.
 Basically, it can be regarded as a modified version of the GWS electroweak theory,
 with the three flavor neutrinos there reduced to one basic neutrino.
 We study a renormalized version of this model
 and find that three types of states in this model
 may be interpreted as three flavor neutrino states.

 Neutrino mass states are associated with low-lying eigenstates of the total Hamiltonian
 within the smallest subspace of the total state space, which contains a basic neutrino state
 and is invariant under the action of the interaction Hamiltonian.
 Based on a first-order perturbation treatment to flavor neutrino states and
 on variational arguments within the above-discussed subspace, 
 we obtain an approximate expression for neutrino mass states.
 Making use of this expression, an estimate is obtained for a ratio of neutrino mass differences,
 which turns out to be about half of the experimental result.
 Moreover, an estimate to the scale of neutrino mass is gotten, 
 which lies in the same order of magnitude as those
 obtained from analyses of cosmological data.
 These results suggest that the proposed model should be worth
 consideration in the search for a satisfactory explanation to neutrino oscillation.

 We list below main approximations used in the derivation of the above-discussed estimates.
\begin{itemize}
  \item  (i)  The  states $|\nu_\kappa\ra$ can be
  approximated by their first-order perturbation expansions [see Eq.(\ref{varphi0-eta-ka})].
  \item  (ii) Equation (\ref{nu-e1}),
 which is obtained under the assumption that the main contribution to the rhs of Eq.(\ref{|nu-e-1>})
 comes from low-lying momentum states,
 gives a good approximation to Eq.(\ref{|nu-e-1>}).
  \item (iii) In the derivation of Eq.(\ref{H-psis-ms}), variations
 in the subspaces $\E^\nu_i$ of $i=2,3,4,5$ are neglected and one term in $\delta f_2$ is not considered.
\end{itemize}

\subsection{Some further discussions}

 Firstly, we discuss a possible direction of improvement in future computation of neutrino mass properties.
 Among the three approximations listed above, the third one seems the most significant.
 It is reasonable to expect that, when contributions from the subspaces $\E^\nu_i$ of $i=2,3,4,5$
 and from the above-mentioned term in $\delta f_2$ are taken into account,
 the expression of neutrino mass states in Eq.(\ref{psi-s}) may be considerably improved.
 In fact, in this equation the three flavor neutrino states have equal weights,
 while, experimental data suggest that the weights should be different. 
 As discussed in the appendix, the above-mentioned term in $\delta f_2$ 
 may require some change in the weights.

 Next, we discuss some impacts of the following prediction of the proposed model,
 that is,  the flavor neutrino states $|\nu_\kappa\ra$ are not orthogonal to each other,
 but have an overlap given by $\la\nu_\kappa|\nu_{\kappa'}\ra
 = \la\varphi_0|\varphi_0\ra$ for $\kappa \ne \kappa'$.
 A condition for this result to be consistent with experimental results has been discussed
 in Sec.\ref{sect-F-neutrino}.

 One impact is that, when only flavor neutrino states are considered in the application of Born's rule,
 the ordinary treatment of making use of von Neumann's projective measurements \cite{Neumann-qm}
 may be invalid.
 This problem is usually not serious, because charged particles are usually involved in experiments of neutrinos,
 the former of which have orthogonal states. 
 The problem can also be solved by making use of some schemes more general than von Neumann's,
 which do not require orthogonality of the generated states.
 For example, one may consider the so-called POVM measurements \cite{nc-book},
 which has been studied extensively in the passed two decades.

 Another impact is related to the three-dimensional mass matrix for neutrino
 discussed in the section of introduction.
 In the proposed model, a corresponding matrix is written as $\la \nu_\kappa|H|\nu_{\kappa'}\ra$,
 which is given in a nonorthogonal basis.

 Finally, we discuss some consequence of the interpretation of 
 the states $|\nu_\kappa\ra$ as flavor neutrino states.
 This interpretation implies that interactions of flavor neutrino states with other particles
 may be more complicated than those described in the SM.
 For example, in the proposed model, possibly 
 the processes $\nu_\kappa + {\kappa'}^+ \to W^+$ with $\kappa \ne \kappa'$ 
 are not completely forbidden, though they are much suppressed
 compared with the processes with $\kappa = \kappa'$ (see discussions given in Sec.\ref{sect-F-neutrino});
 in contrast, the former processes are completely forbidden in the SM.

 We note that the above prediction is not necessarily in confliction with experimental results.
 In fact, as reported in LSND \cite{LSND1,LSND2} and MiniBooNE \cite{mini1,mini2}
 experiments, there is $>3\sigma $ evidence of $\ov\nu_e$ appearance in a
 $\ov\nu_\mu$ beam, and a $\nu_e$ signal appeared in a $\nu_\mu$ beam in MiniBooNE.
 These experiments may be explained, if processes like $\ov\nu_\mu + e \to W^-$ may happen
 with a small probability, which implies that with a small probability a flavor neutrino $\ov\nu_\mu$ may 
 behave like a flavor neutrino $\ov\nu_e$.

 In future investigations, it would be of interest to study whether the proposed model could be
 really useful in explaining the above-discussed anomaly observed in neutrino experiments.
 One may also study other anomalies, 
 particularly, the so-called reactor anomaly \cite{reactor1,reactor2,reactor3},
 the Gallium anomaly \cite{G-anomaly1,G-anomaly2,G-anomaly3,G-anomaly4},
 and an anomaly observed in Nutev experiments \cite{Nutev}.

\acknowledgments

 The author is grateful to Guijun Ding for valuable discussions and suggestions.
 This work was partially supported by the National Natural Science Foundation of China under Grant
 Nos.~11535011 and 11775210.\\

\appendix

\section{A possibility of reducing $\delta f_2$}\label{app-f2}

 In this appendix, we discuss a possibility of further reducing $\delta f_2$
 for variations within the subspace $\E^\nu_\kappa$ for which $c_\kappa^{(s)}=-1$.
 To this end, one may consider a modification to $|\psi_s\ra$, denoted by $|\psi_s'\ra$,
\begin{gather}\label{psi-s'}
 |\psi_s'\ra = |\varphi_0\ra + \sum_\kappa c_\kappa^{(s)} (|\eta_\kappa\ra +  |\xi_{\kappa}\ra),
\end{gather}
 where $|\xi_{\kappa}\ra =0$ if $c_\kappa^{(s)}=1$, and $|\xi_{\kappa}\ra \ne 0$ if $c_\kappa^{(s)}=-1$.
 Under the same variation $|\delta\psi\ra$ as that discussed in the main text,
 $\la\delta\psi|H|\psi_s'\ra$ gets the following additional terms,
\begin{gather*}
 c_{\kappa}  \big(\la\delta \varphi_0|  V_{\kappa} |\xi_{\kappa}\ra
 + \la\delta\eta_{\kappa}| H^{\nu}_{Z} |\xi_{\kappa}\ra
  + \la\delta\eta_{\kappa}|  V_{\kappa} |\xi_{\kappa}\ra \big),
\end{gather*}
 and $\lambda \la \delta \psi|\psi\ra$ gets
 an additional term of $ -  c_\kappa \lambda \la \delta\eta_\kappa|\xi_{\kappa(s)}\ra $,
 where $\kappa$ is determined by $s$ according to the relation $c_\kappa^{(s)}=-1$.
 These additional terms can be canceled, together with $\delta f_2$ in Eq.(\ref{delta-f2}) and the term
 $ \la\delta \varphi_0|(E_{\nu_\kappa}^0 -\lambda) |\varphi_0\ra$ in $\delta f_1$, if
 $|\xi_{\kappa}\ra$ may be chosen such that
\begin{gather}
 P_\nu^Z V_\kappa|\xi_{\kappa}\ra + (E_{\nu_\kappa}^0 -\lambda) |\varphi_0\ra= 0,
 \\   H^{\nu}_{Z}|\xi_{\kappa}\ra +  V_{\kappa} |\xi_{\kappa}\ra - \lambda |\xi_{\kappa}\ra = 2 V_{\kappa} |\varphi_0\ra,
\end{gather}
 where $P_\nu^Z$ represents the projection operator for the subspace $\E_\nu^Z$.
 Then, $\delta F$ remains small for what ever variation from $|\psi_s'\ra$ within the space $\E^\nu_\kappa$.

\end{document}